\documentclass[reprint,pra,amssymb,amsmath,superscriptaddress,
longbibliography]{revtex4-1}

\pdfoutput=1

\usepackage[colorlinks, citecolor=blue, linkcolor=blue,
urlcolor=blue]{hyperref}
\usepackage{longtable}
\usepackage{amsthm}
\newtheorem{lemma}{Lemma}

\usepackage{mathptmx}
\usepackage{mathrsfs}
\usepackage{bbold}
\usepackage[LY1]{fontenc}

\newcommand\ket[1]{\,|#1\rangle}
\newcommand\bra[1]{\langle #1|}

\newcommand\Span[0]{\operatorname{Span}}
\renewcommand{\phi}{\varphi}

\newcommand\bket[1]{|\mathfrak{#1}\rangle}

\newcommand\bitz[0]{\mathfrak{0}}
\newcommand\bito[0]{\mathfrak{1}}
\newcommand\sA[0]{\mathrm{A}}
\newcommand\sB[0]{\mathrm{B}}
\newcommand\sE[0]{\mathrm{E}}
\newcommand\sx[0]{\mathrm{x}}
\newcommand\sxB[0]{\mathrm{x,B}}
\newcommand\sInit[0]{\mathrm{init}}
\newcommand\sCTRL[0]{\mathrm{CTRL}}
\newcommand\sSWAPOL[0]{\mathrm{S\text{-}01}}
\newcommand\sSWAPLO[0]{\mathrm{S\text{-}10}}
\newcommand\sSWAPALL[0]{\mathrm{S\text{-}ALL}}

\begin{document}
\title{Experimentally feasible protocol for
semiquantum key distribution}
\author{Michel Boyer}
\email{boyer@iro.umontreal.ca}
\affiliation{D\'epartement IRO, Universit\'e de Montr\'eal,
Montr\'eal (Qu\'ebec) H3C 3J7, Canada}
\author{Matty Katz}
\author{Rotem Liss}
\email{rotemliss@cs.technion.ac.il}
\author{Tal Mor}
\email{talmo@cs.technion.ac.il}
\affiliation{Computer Science Department, Technion, Haifa 3200003,
Israel}

\begin{abstract}
Quantum key distribution (QKD) protocols make it possible for
two quantum parties to generate a secret shared key.
Semiquantum key distribution (SQKD) protocols, such as ``QKD with
classical Bob'' and ``QKD with classical Alice'' (that have both been
proven robust), achieve this goal even if one of the parties is
classical. However, existing SQKD protocols are
not experimentally feasible with current technology.
Here we suggest a new protocol, ``Classical Alice with
a controllable mirror'', that can be experimentally implemented
with current technology (using $4$-level systems instead of qubits), 
and we prove it to be robust.
\end{abstract}

\pacs{03.67.Dd, 03.65.Sq}

\maketitle

\section{\label{sec_intro}Introduction}
Quantum key distribution (QKD) makes it possible for two legitimate
parties, Alice and Bob, to generate an information-theoretically
secure key~\cite{bb84}, that is secure against any possible attack
(of the adversary Eve) allowed by the laws of quantum physics.
Alice and Bob use an insecure quantum channel and an authenticated
classical channel.

Semiquantum key distribution (SQKD) protocols limit one of the parties
to be classical, while still giving a secure key~\cite{cbob07}.
As explained in~\cite{cbob07,sqkd09}, such SQKD protocols are
interesting from both the conceptual and the practical points of view;
moreover, in a network of one quantum center and many
classical ``users'', the classical users may even be oblivious
to being involved in a quantum cryptographic protocol.

The first SQKD protocol was ``QKD with classical Bob''~\cite{cbob07};
later, the ``QKD with classical Alice''~\cite{calice09,calice09comment}
protocol was suggested, as well as various other SQKD protocols
(see for example~\cite{LC2008,SDL2013,YYLH2014}).
Most of the SQKD protocols have been proven ``robust'':
namely~\cite{cbob07}, any successful attack by an adversary necessarily
induces some noise that the legitimate parties may notice.
A few of them also have their full security
analyzed~\cite{cbob_security15}.

However, to the best of our knowledge, all the currently existing SQKD
protocols cannot be experimentally constructed in a secure way
by using current technology, because, as explained below, one of the
``classical'' operations (SIFT) cannot be securely implemented.

We present a new and feasible SQKD protocol that can be
experimentally constructed by using a ``controllable mirror''.
It is based on ``QKD with classical
Alice''~\cite{calice09,calice09comment},
but it is slightly more complicated,
because it uses $4$-level systems instead of qubits ($2$-level systems),
and because it requires Alice to choose one of four operations
(instead of two). We prove this protocol to be robust.

In Sec.~\ref{sec_calice} we present the original
``QKD with classical Alice'' protocol,
and in Sec.~\ref{sec_infeasible} we explain why this protocol
and the other currently existing protocols are
experimentally infeasible with current technology.
In Sec.~\ref{sec_mirror_protocol} we present
the ``Classical Alice with a controllable mirror'' SQKD protocol,
and in Sec.~\ref{sec_robustness} we prove it to be robust.
We conclude in Sec.~\ref{sec_conclusion}.

\section{\label{sec_calice}Quantum key distribution
with classical Alice}
In the ``QKD with classical Alice'' protocol~\cite{calice09}
(the name is following~\cite{calice09comment}), in each round,
the originator Bob sends to Alice the qubit state $\ket{\text{+}}$.
Then, Alice randomly chooses one of two classical operations:
CTRL --- reflect the qubit to Bob, or SIFT --- measure the qubit
in the computational (i.e., the classical) basis
$\{\bket{0}, \bket{1}\}$ and resend it to Bob.
Bob then measures the qubit he receives from classical Alice,
choosing randomly the measurement basis (the computational basis
or the Hadamard basis $\{\ket{\text{+}}, \ket{-}\}$).
After $N$ qubits have been sent and received,
Alice publicly announces her choice (CTRL or SIFT) for each round,
and Bob publicly announces his basis choice for each round.
Then, Alice and Bob check the error rates in the CTRL bits
and in a random subset of the SIFT bits,
aborting if they are too high.
Finally, Alice and Bob perform error correction
and privacy amplification on the remaining SIFT bits measured by Bob
in the computational basis, so that they get a final identical key
that is completely secret.

As proven in~\cite{calice09comment},
``QKD with classical Alice''~\cite{calice09} is completely
robust against eavesdropping. The proof of robustness was extended
in~\cite{calice_photonic10_article}
to include photonic implementations and multi-photon pulses.

\section{\label{sec_infeasible}The experimental infeasibility
of the SIFT operation in SQKD protocols}
In the SQKD protocols (e.g.,~\cite{cbob07,calice09}),
one of the ``classical'' operations is SIFT:
measuring in the computational basis $\{\bket{0}, \bket{1}\}$
and then resending.
In practical (photonic) implementations,
and especially if limited to the existing technology,
the SIFT operation is very hard to securely implement,
because the generated photon will probably be
at a different timing or frequency,
thus leaking information to the eavesdropper;
see details in~\cite{cbob07comment} (which is a comment
on~\cite{cbob07}) and in the reply~\cite{cbob07comment_reply}.

For example, let us look at the ``QKD with classical Alice'' protocol
implemented with two {\em classical} modes, $\bket{0}$ and $\bket{1}$,
describing two pulses (two distinct time-bins) on a single arm.
The photon can be either in one pulse, in the other,
or in a superposition (a non-classical state).
In this case, the SIFT operation requires Alice to measure
the two pulses, to generate a single photon in a state depending
on the measurement outcome, and to resend it to Bob;
while Alice can implement the CTRL operation
simply by using a mirror (reflecting both pulses).
In this case, it is indeed very difficult for Alice
to regenerate the SIFT photon exactly at the right timing,
so that it is indistinguishable from a CTRL photon.

Furthermore, in~\cite{cbob07comment} it was shown that even if Alice
could (somehow) have the machinery to perform SIFT with perfect timing,
Eve would still be able to attack the protocol
by taking advantage of the fact that Alice's detectors are imperfect:
Eve's attack is modifying the {\em frequency} of the photon generated
by Bob. Alice does not notice the change in frequency.
If Alice performs SIFT,
the photon she generates is in the original frequency;
if she performs CTRL,
the photon she reflects is in the frequency modified by Eve.
Therefore, if Eve is powerful enough, she can measure the frequency
and tell whether Alice used SIFT or CTRL.
If Eve finds out that Alice used SIFT, she can copy the bit
sent by Alice in the computational basis;
if she finds out that Alice used CTRL,
she shifts the frequency back to the original frequency.
(A very similar attack works for other implementations, too ---
e.g., for polarization-based or phase-based implementations.)
This ``tagging'' attack makes it possible for Eve to get
full information on the key without inducing noise.

\section{\label{sec_mirror_protocol}The controllable mirror protocol
for QKD with classical Alice}
We suggest a new SQKD protocol, similar to
``QKD with classical Alice'', that is experimentally feasible:
in the original protocol of ``QKD with classical Alice'', Alice could
choose only between two operations (CTRL and SIFT);
in our new protocol, Alice may choose between four operations
(CTRL, SWAP-10, SWAP-01, and SWAP-ALL).
This protocol avoids the need of using the infeasible operation SIFT.
The two operations SWAP-10 and SWAP-01 correspond to two 
possible reflections of {\em pulses}
by using a controllable mirror. Those operations cannot be described
by qubit notations, so below we use $4$-level system notations.
Our new protocol is based on the Fock space notations:
in those notations, the state $\ket{m_\bito, m_\bitz}$ represents
$m_\bito$ indistinguishable photons in
the mode of the qubit-state $\bket{1}$
and $m_\bitz$ indistinguishable photons in
the mode of the qubit-state $\bket{0}$.
More details about the Fock space notations are given
in Appendix~\ref{appendix_fock}.

This protocol is experimentally feasible and is safe against the
``tagging'' attack described in~\cite{cbob07comment}.
Moreover, we prove this protocol to be completely robust
against an attacker Eve that can
do anything allowed by the laws of quantum physics,
including the possibility of sending multi-photon pulses (namely,
assuming that Eve may use any quantum state consisting of the
two modes $\bket{0}$ and $\bket{1}$: that is, any superposition of
the Fock states $\ket{m_\bito, m_\bitz}$).

We can describe the new protocol in terms of photon pulses that
correspond to two distinct time-bins,
and of a controllable mirror operated by Alice:
in this case, the CTRL operation corresponds
to operating the mirror on both pulses (reflecting both pulses back
to the originator, Bob); 
the SWAP-10 operation corresponds
to operating the mirror only on the $\bket{0}$ pulse
while measuring the other pulse
(and similarly for the SWAP-01 operation and the $\bket{1}$ pulse);
and the SWAP-ALL operation corresponds
to measuring all the pulses, without reflecting any of them.

For the experimental implementation, we note that
a (very slow) mechanically-moved mirror is trivial to implement;
a much faster device can be electronically implemented by using
standard optical elements (that are commonly used in QKD):
a Pockels cell (that can change the polarization of the photon(s)
in one of the pulses) and a polarizing beam splitter
(that makes it possible to split the two different pulses into
two paths, because they are now differently polarized).
Like other (fast) QKD experimental settings,
implementation is feasible but is not trivial.

Let Alice's initial probe be in the vacuum state $\ket{0,0}_\sA$,
and let us assume that a single photon is arriving from Bob;
thus, the system {\em as a whole} can be described as
a $4$-level system (a single photon in four modes).
Alice's operations are as follows:
\begin{description}
\item[$\mathbf{I}$ (CTRL)] Do nothing:
\begin{equation}
I \ket{0,0}_\sA \ket{m_\bito,m_\bitz}_\sB
= \ket{0,0}_\sA \ket{m_\bito,m_\bitz}_\sB
\end{equation}
\item[$\mathbf{S_1}$ (SWAP-10)] Swap half of Alice's probe
(the left mode) with the $\ket{m_\bito}_\sB$ half of Bob's state:
\begin{equation}
S_1 \ket{0,0}_\sA \ket{m_\bito,m_\bitz}_\sB
= \ket{{m_\bito,0}}_\sA \ket{0,m_\bitz}_\sB
\end{equation}
\item[$\mathbf{S_0}$ (SWAP-01)] Swap half of Alice's probe
(the right mode) with the $\ket{m_\bitz}_\sB$ half of Bob's state:
\begin{equation}
S_0 \ket{0,0}_\sA \ket{m_\bito,m_\bitz}_\sB
= \ket{0,m_\bitz}_\sA \ket{m_\bito,0}_\sB
\end{equation}
\item[$\mathbf{S}$ (SWAP-ALL)] Swap the entire probe of Alice with
the entire state $\ket{m_\bito,m_\bitz}_\sB$ of Bob:
\begin{equation}
S \ket{0,0}_\sA \ket{m_\bito,m_\bitz}_\sB
= \ket{m_\bito,m_\bitz}_\sA \ket{0,0}_\sB
\end{equation}
\end{description}
After each of the three SWAP operations,
Alice measures her probe (the $\ket{\cdot}_\sA$ state)
in the computational basis and sends to Bob the $\ket{\cdot}_\sB$ state.
If there is no noise and no eavesdropping,
and if we analyze the ``ideal case''
(in which exactly one photon is arriving from Bob to Alice),
then each round is described by the four-dimensional Hilbert space
$\Span\{\ket{0,0}_\sA \ket{0,1}_\sB, \ket{0,0}_\sA \ket{1,0}_\sB,
\ket{0,1}_\sA \ket{0,0}_\sB, \ket{1,0}_\sA \ket{0,0}_\sB\}$ ---
namely, by a four-level system;
for our protocol, we use this four-level system
instead of the qubit used by BB84 and by many other QKD schemes.
In the most general ``theoretical attack''
(the attack analyzed by standard QKD security proofs),
Eve attacks Alice's and Bob's states using any probe of her choice,
but she cannot modify the four-dimensional Hilbert space
of the protocol: she can only use those four levels.
However, in practical attacks (as analyzed in our robustness analysis),
Eve may use an extended Hilbert space (the entire Fock space).

While Eve is fully powerful, it is common to assume that Alice and Bob
are limited to use only current technology.
In particular, Alice and Bob are limited in the sense
that they cannot {\em count} the number of photons in each mode,   
but can only check whether a detector corresponding
to a specific mode clicks (detects at least one photon in that mode)
or not (detects an empty mode).
For our protocol to be practical (and for our robustness analysis
to be stronger),
we assume that Alice and Bob are indeed limited in that sense.
Therefore, when Alice and Bob measure in the computational basis,
their measurement results are
denoted as $\hat{m}_\bito \hat{m}_\bitz$, with
$\hat{m}_\bitz, \hat{m}_\bito \in \{0, 1\}$.
Similarly, when Bob measures in the Hadamard basis,
his measurement result is $\hat{m}_- \hat{m}_+$, with
$\hat{m}_+, \hat{m}_- \in \{0, 1\}$.

This limitation leads to the definition of ``sum'', as follows:
let us look at a measurement result of Alice or Bob (that is $00$,
$01$, $10$, or $11$). The ``{\em sum}'' of this
measurement result is the number of distinct modes
detected to be non-empty during the measurement
(namely, the sum of the digits in the measurement result).
This definition is summarized in Table~\ref{calice_mirror_measure_defs}.

\begin{table}
\caption{\label{calice_mirror_measure_defs} The four possible
measurement results by Alice or Bob
(measuring in the computational basis),
depending on the state obtained by him or her
(that is represented in the Fock space notations).}
\begin{center}
\begin{tabular}{lll}
\hline\hline
Obtained State & Measurement Result & ``Sum'' \\
\hline
$\ket{0,0}$ & $00$ & $0$ \\
$\ket{0,m_\bitz}$ ($m_\bitz > 0$) & $01$ & $1$ \\
$\ket{m_\bito,0}$ ($m_\bito > 0$) & $10$ & $1$ \\
$\ket{m_\bito,m_\bitz}$ ($m_\bito > 0, m_\bitz > 0$)
& $11$ & $2$ \\ \hline\hline
\end{tabular}
\end{center}
\end{table}

The protocol consists of the following steps:
\begin{enumerate}
\item \label{calice_mirror_round} In each of the $N$ rounds,
Bob sends to Alice the state $\ket{\text{+}}_\sB$;
Alice randomly chooses one of her four classical operations
(CTRL, SWAP-10, SWAP-01, or SWAP-ALL) and sends the result back to Bob;
and Bob measures the state he receives,
choosing randomly whether to measure in the computational basis
or in the Hadamard basis.
\item Alice reveals her operation choices
(CTRL, SWAP-x ($x \in \{01,10\}$), or SWAP-ALL;
Alice does \underline{not} reveal her choices between
SWAP-10 and SWAP-01, that she keeps as a secret bit string),
and Bob reveals his basis choices.
They discard all CTRL bits Bob measured in the computational basis
and all SWAP-x bits he measured in the Hadamard basis.
\item For each of the SWAP-x and SWAP-ALL states,
Alice and Bob reveal the ``sums'' of their measurement results.
\item \label{calice_mirror_interpretation} Alice and Bob interpret
their measurement results:
they consider several types of measurement results as errors, losses,
or valid results. See Tables~\ref{robust_analysis_interp_I}-
\ref{robust_analysis_interp_S} for the details.
\item \label{calice_mirror_bits} For all the SWAP-x ($x \in \{01,10\}$)
states, if Bob's ``sum'' is $1$ and Alice's ``sum'' is $0$,
then Alice and Bob share a (secret) bit $b$,
because Alice knows (in secret) what operation $S_{1-b}$ she performed,
and Bob knows (in secret) what mode $\ket{b}$ he detected.
Each one of Alice and Bob keeps this sequence of bits $b$
as his or her bit string.
\item Alice and Bob reveal some random subset of their bit strings,
compare them, and estimate the error rate
(this is the error rate in the way from Alice back to Bob).
They abort the protocol if the error rate in those bits, or
any of the error rates measured in
Step~\ref{calice_mirror_interpretation}, is above a specified threshold.
They discard the revealed bits.
\item Alice and Bob perform error correction and privacy amplification
processes on the remaining bit string,
yielding a final key that is identical for Alice and Bob
and is fully secure from any eavesdropper.
\end{enumerate}

\begin{table}
\caption{\label{robust_analysis_interp_I} Interpretations of
Bob's measurement results for CTRL states.}
\begin{center}
\begin{tabular}{ll}
\hline\hline
Bob's Result & Interpretation \\
\hline
$00$ & a loss \\
$01$ (i.e., $\ket{\text{+}}$) & a legal result \\
$10$ (i.e., $\ket{-}$) & an error \\
$11$ & an error \\ \hline\hline
\end{tabular}
\end{center}
\end{table}

\begin{table}
\caption{\label{robust_analysis_interp_Sx} Interpretations of
Alice's and Bob's measurement results for SWAP-x states.}
\begin{center}
\begin{tabular}{lll}
\hline\hline
Alice's ``Sum'' & Bob's ``Sum'' & Interpretation \\
\hline
$0$ & $0$ & a loss \\
$0$ & $1$ & Alice and Bob share a bit \\
$1$ & $0$ & Alice and Bob do not share a bit \\
$1$ & $1$ & an error \\
$0$ or $1$ & $2$ & an error \\
$2$ & & impossible \\ \hline\hline
\end{tabular}
\end{center}
\end{table}

\begin{table}
\caption{\label{robust_analysis_interp_S} Interpretations of
Alice's and Bob's measurement results for SWAP-ALL states.}
\begin{center}
\begin{tabular}{lll}
\hline\hline
Alice's Result & Bob's Result & Interpretation \\
\hline
$00$ & $00$ & a loss \\
$01$ or $10$ & $00$ & a legal result \\
$11$ & $00$ & an error \\
& $01$, $10$, or $11$ & an error \\ \hline\hline
\end{tabular}
\end{center}
\end{table}

Notice that Bob does not have a special role in the beginning:
he always generates the same state, $\ket{\text{+}}$.
It is even possible that the adversary Eve generates
this state instead of him.

\section{\label{sec_robustness}Robustness analysis}
To prove robustness, we will prove that for Eve's attack to be
undetectable by Alice and Bob (namely, for Eve's attack not to cause
any errors), it must not give Eve any information.

Eve's attack on a state can be performed in both directions:
from the source (Bob) to Alice, Eve applies $U$;
from Alice back to Bob, Eve applies $V$.
We may assume, without limiting generality, that Eve uses
a fixed probe space $\mathcal{H}_\sE$ for her attacks.

According to the definition of robustness, we will prove that if,
during a run of the protocol, no error can be detected
by Alice and Bob, then Eve gets no information on the raw key.

If Alice and Bob cannot find any error, the following
conditions must be satisfied for all the measurement results that
were not discarded due to basis mismatch:
\begin{enumerate}
\item For all CTRL states, Bob's measurement result
(in the Hadamard basis) must not be $10$ or $11$
(namely, Bob must never detect any photon in the
$\ket{-}$ mode).
\label{cond_ctrl}
\item For all SWAP-x states, Alice's ``sum'' and Bob's ``sum''
(in the computational basis) must not be both $1$. \label{cond_sift_11}
\item For all SWAP-x states, Bob's ``sum''
(in the computational basis) must not be $2$
(namely, Bob's measurement result must not be $11$).
\label{cond_sift_2}
\item For all SWAP-x states, no error (that may be detected
during the protocol) can exist. In other words:
\label{cond_sift_test}
\begin{enumerate}
\item For all SWAP-10 states, Bob's measurement result
(in the computational basis) must not be $10$. \label{cond_sift_test1}
\item For all SWAP-01 states, Bob's measurement result
(in the computational basis) must not be $01$. \label{cond_sift_test0}
\end{enumerate}
\item For all SWAP-ALL states, Alice's measurement result
must not be $11$. \label{cond_sift_all_alice}
\item For all SWAP-ALL states, Bob's measurement result
must not be $01$, $10$, or $11$. \label{cond_sift_all_bob}
\end{enumerate}

We now analyze each round of the protocol.
After the round begins, the source (Bob) sends to Alice the state
$\ket{0,1}_\sxB \in \mathcal{H}_\sB$.
Eve can now interfere: she attaches her own probe state
(in the Hilbert space $\mathcal{H}_\sE$) and applies
the unitary transformation $U$. The resulting Bob+Eve state
(including Eve's probe) is of the form
\begin{equation}
\ket{\psi_\sInit} \triangleq
\sum_{m_\bito, m_\bitz} \ket{m_\bito, m_\bitz}_\sB
\ket{E_{m_\bito, m_\bitz}}_\sE,
\end{equation}
where $\ket{E_{i,j}}_\sE$ are non-normalized vectors
in $\mathcal{H}_\sE$.

Condition~\ref{cond_sift_all_alice} means that
$\ket{E_{m_\bito,m_\bitz}}_\sE = 0$ for all $m_\bito, m_\bitz$
satisfying $m_\bito > 0$ and $m_\bitz > 0$. Therefore,
\begin{equation}
\ket{\psi_\sInit} = \ket{\phi_{1,0}} + \ket{\phi_{0,1}}
+ \ket{\phi_{0,0}},
\end{equation}
with
\begin{eqnarray}
\ket{\phi_{1,0}} &\triangleq& \sum_{m_\bito > 0} \ket{m_\bito,0}_\sB
\ket{E_{m_\bito,0}}_\sE, \label{define_psi10} \\
\ket{\phi_{0,1}} &\triangleq& \sum_{m_\bitz > 0} \ket{0,m_\bitz}_\sB
\ket{E_{0,m_\bitz}}_\sE, \label{define_psi01} \\
\ket{\phi_{0,0}} &\triangleq& \ket{0,0}_\sB \ket{E_{0,0}}_\sE.
\label{define_psi00}
\end{eqnarray}

Alice now applies one of the four possible operations
(CTRL = $I$, SWAP-10 = $S_1$, SWAP-01 = $S_0$, or SWAP-ALL = $S$)
and destructively measures her probe state.
The (non-normalized) state of the Bob+Eve system
after Alice's operation (and measurement)
is written in Table~\ref{robust_proof_state}.

\begin{table}
\caption{\label{robust_proof_state} The (non-normalized) state
of the Bob+Eve system after Alice's operation, given Alice's ``sum''.
Note that the states $\ket{\phi_{1,0}}$, $\ket{\phi_{0,1}}$, and
$\ket{\phi_{0,0}}$ are defined in
\eqref{define_psi10}-\eqref{define_psi00}.}
\begin{center}
\begin{tabular}{lllll}
\hline\hline
Operation & & Alice & & Bob+Eve State \\
\hline
CTRL & & & & $\ket{\psi_\sCTRL} \triangleq \ket{\phi_{1,0}} +
\ket{\phi_{0,1}} + \ket{\phi_{0,0}}$ \\ \hline
SWAP-10 & & $0$ & & $\ket{\psi_\sSWAPLO^{(0)}}
\triangleq \ket{\phi_{0,1}} +
\ket{\phi_{0,0}}$ \\
SWAP-01 & & $0$ & & $\ket{\psi_\sSWAPOL^{(0)}}
\triangleq \ket{\phi_{1,0}} +
\ket{\phi_{0,0}}$ \\ \hline
SWAP-10 & & $1$ & & $\rho_\sSWAPLO^{(1)}
\triangleq \sum_{m_\bito > 0} \ket{0,0}_\sB \bra{0,0}
\otimes \ket{E_{m_\bito,0}}_\sE \bra{E_{m_\bito,0}}$ \\
SWAP-01 & & $1$ & & $\rho_\sSWAPOL^{(1)}
\triangleq \sum_{m_\bitz > 0} \ket{0,0}_\sB \bra{0,0}
\otimes \ket{E_{0,m_\bitz}}_\sE \bra{E_{0,m_\bitz}}$
\\ \hline
SWAP-ALL & & & & $\rho_\sSWAPALL \triangleq \rho_\sSWAPLO^{(1)}
+ \rho_\sSWAPOL^{(1)} + \ket{\phi_{0,0}} \bra{\phi_{0,0}}$
\\ \hline\hline
\end{tabular}
\end{center}
\end{table}

Then, Eve applies a second unitary transformation $V$ on the state
sent from Alice to Bob (and on her own probe state).
According to conditions~\ref{cond_sift_11},
\ref{cond_sift_2}, and~\ref{cond_sift_all_bob},
the density matrices $V \rho_\sSWAPLO^{(1)} V^\dagger$,
$V \rho_\sSWAPOL^{(1)} V^\dagger$, and $V \rho_\sSWAPALL V^\dagger$
must only overlap with $\ket{0,0}_\sB$. It follows that there
exists $\ket{H_{0,0}}_\sE \in \mathcal{H}_\sE$ such that
\begin{equation}\label{V00}
V \ket{\phi_{0,0}} = \ket{0,0}_\sB \ket{H_{0,0}}_\sE.
\end{equation}

Let $V \ket{\phi_{1,0}} = \sum_{m_\bito,m_\bitz}
\ket{m_\bito,m_\bitz}_\sB \ket{F_{m_\bito,m_\bitz}}_\sE$.
Let us look at a SWAP-01 state for which Alice's ``sum'' is $0$.
For this state, the Bob+Eve state after Eve's attack is
\begin{eqnarray}
V \ket{\psi_\sSWAPOL^{(0)}}
&=& V \ket{\phi_{1,0}} + V \ket{\phi_{0,0}} \\
&=& \sum_{m_\bito,m_\bitz} \ket{m_\bito,m_\bitz}_\sB
\ket{F_{m_\bito,m_\bitz}}_\sE + \ket{0,0}_\sB
\ket{H_{0,0}}_\sE,
\nonumber
\end{eqnarray}
and following conditions~\ref{cond_sift_test0} and~\ref{cond_sift_2},
Bob must not detect a photon in the $\bket{0}$ mode
(otherwise, the error may be detected during the protocol).
Therefore,
$\ket{F_{m_\bito,m_\bitz}}_\sE = 0$ for all $m_\bitz > 0$.
It follows that
\begin{equation}
V \ket{\phi_{1,0}} = \sum_{m_\bito > 0}
\ket{m_\bito,0}_\sB \ket{F_{m_\bito,0}}_\sE
+ \ket{0,0}_\sB \ket{F_{0,0}}_\sE. \label{V10}
\end{equation}
Similarly (following
conditions~\ref{cond_sift_test1} and~\ref{cond_sift_2}),
\begin{equation}
V \ket{\phi_{0,1}} = \sum_{m_\bitz > 0}
\ket{0,m_\bitz}_\sB \ket{G_{0,m_\bitz}}_\sE
+ \ket{0,0}_\sB \ket{G_{0,0}}_\sE. \label{V01}
\end{equation}

Now, \eqref{V00}, \eqref{V10}, and \eqref{V01}
imply that if Alice applies CTRL,
the Bob+Eve state after Eve's attack is
\begin{equation}
V \ket{\psi_\sCTRL} = \sum_{m > 0}
\left[ \ket{m,0}_\sB \ket{F_{m,0}}_\sE
+ \ket{0,m}_\sB \ket{G_{0,m}}_\sE \right] + \ket{0,0}_\sB \ket{H}_\sE
\end{equation}
with $\ket{H}_\sE = \ket{F_{0,0}}_\sE + \ket{G_{0,0}}_\sE
+ \ket{H_{0,0}}_\sE$.
Following condition~\ref{cond_ctrl}, the probability of
Bob getting a photon in the $\ket{-}$ mode must be $0$.

We now use the following Lemma,
the proof of which is given in Appendix~\ref{appendix_lemma_proof}:
\begin{lemma}\label{lemma11}
If $\ket{\psi'} = \sum_{m > 0} \left[ \ket{m,0}_\sB \ket{F_{m,0}}_\sE
+ \ket{0,m}_\sB \ket{G_{0,m}}_\sE \right] + \ket{0,0}_\sB \ket{H}_\sE$
is a bipartite state in $\mathcal{H}_\sB \otimes \mathcal{H}_\sE$,
and if there is a zero probability that Bob gets a photon
in the $\ket{-}$ mode,
then $\ket{F_{1,0}}_\sE = \ket{G_{0,1}}_\sE$,
and $\ket{F_{m,0}}_\sE = \ket{G_{0,m}}_\sE = 0$ for all $m > 1$.
\end{lemma}

Applying Lemma~\ref{lemma11}, we deduce that
$\ket{F_{m,0}}_\sE = \ket{G_{0,m}}_\sE = 0$ for all $m > 1$, and that
$\ket{F_{1,0}}_\sE = \ket{G_{0,1}}_\sE \triangleq \ket{F}_\sE$.

It follows that the Bob+Eve states after Eve's attack,
when Alice performed SWAP-x and her ``sum'' is $0$ (those are
the only states for which Alice and Bob may share a bit), are:
\begin{eqnarray}
V \ket{\psi_\sSWAPLO^{(0)}} &=& \ket{0,1}_\sB \ket{F}_\sE
+ \ket{0,0}_\sB \left[ \ket{G_{0,0}}_\sE + \ket{H_{0,0}}_\sE \right] \\
V \ket{\psi_\sSWAPOL^{(0)}} &=& \ket{1,0}_\sB \ket{F}_\sE
+ \ket{0,0}_\sB \left[ \ket{F_{0,0}}_\sE + \ket{H_{0,0}}_\sE \right]
\end{eqnarray}
Therefore, the state of Eve's probe is independent of all
Alice's and Bob's shared bits, and is equal to $\ket{F}_\sE$
whenever Alice and Bob share a bit. Eve can thus get no information
on the bits shared by Alice and Bob without being detected.

\section{\label{sec_conclusion}Conclusion}
We have presented a new semiquantum key distribution protocol,
and have proved it robust (security analysis is left for the future).
Unlike all the previous SQKD protocols, our new protocol
can be experimentally implemented in a secure way.

In this paper, we have suggested a solution for
a practical security problem of SQKD protocols,
that was discussed in Sec.~\ref{sec_infeasible}
and in~\cite{cbob07comment}.
We note that QKD protocols have, too, some security weaknesses
in their practical implementations, such as
the ``Photon-Number Splitting'' attack~\cite{BLMS00},
the ``Bright Illumination'' attack~\cite{makarov10},
the ``Fixed Apparatus'' attack~\cite{fixed_apparatus14},
and other practical attacks.
While some of those security weaknesses can be mitigated,
full security proofs for practical implementations
are still out of reach.
A future extension of this paper may check
to what extent the practical implementations of
the SQKD protocols discussed in this paper
suffer from the same practical security problems as common QKD
protocols, and whether insights from SQKD protocols
(and the methods described in this paper)
may help in solving practical security
problems of both SQKD and fully-quantum QKD protocols.

\begin{acknowledgments}
The work of TM and RL was partly supported
by the Israeli MOD Research and Technology Unit,
and by the Gerald Schwartz \& Heather Reisman Foundation.
\end{acknowledgments}

\appendix

\section{\label{appendix_fock}Fock space notations}
The Fock space notations, that serve as an extension of the qubit space,
are defined as follows:
the Fock basis vector $\ket{0,1}$ represents
a single photon in the $\bket{0}$ state,
and the Fock basis vector $\ket{1,0}$ represents
a single photon in the $\bket{1}$ state.
The vectors $\ket{0,1}$ and $\ket{1,0}$ could, for example,
be two polarization modes,
two arm modes (e.g., arms entering an interferometer),
or two time-bin modes on a single arm.
The qubit space (representing a single photon in one of the two modes)
can be extended to the entire $2$-mode Fock space
\begin{equation}
\mathcal{F} = \Span\{\ket{m_\bito, m_\bitz} \mid
\ m_\bito\geq 0, m_\bitz\geq 0\},
\end{equation}
where the state $\ket{m_\bito, m_\bitz}$ represents
$m_\bito$ indistinguishable photons in
the mode of the qubit-state $\bket{1}$
and $m_\bitz$ indistinguishable photons in
the mode of the qubit-state $\bket{0}$.
In particular, the state $\ket{0,0} \in \mathcal{F}$ is used
for describing absence of photons in both modes (the ``vacuum state'').

Similarly, a single photon in the $\ket{\text{+}}$ mode may
be written as $\ket{0,1}_\sx$
(and similarly for $\ket{-}$ and $\ket{1,0}_\sx$),
and the entire $2$-mode Fock space can be represented as
\begin{equation}
\mathcal{F} = \Span\{\ket{m_-, m_+}_\sx \mid
\ m_-\geq 0, m_+\geq 0\},
\end{equation}
where the state $\ket{m_-, m_+}_\sx$ represents
$m_-$ indistinguishable photons in
the mode of the qubit-state $\ket{-}$
and $m_+$ indistinguishable photons in
the mode of the qubit-state $\ket{\text{+}}$.

In this paper, we shorten the term
``the mode of the qubit-state $\bket{0}$''
to ``the $\bket{0}$ mode'';
and similarly for $\bket{1}, \ket{\text{+}}, \ket{-}$.

\section{\label{appendix_lemma_proof}Proof for Lemma~\ref{lemma11}}
\begin{proof}
If there is a zero probability that Bob gets a photon
in the $\ket{-}$ mode,
then there is a zero probability of measuring any basis state
$\ket{m_-,m_+}_\sxB$ of $\mathcal{H}_\sB$ with $m_- > 0$.

For $m = 1$, since
$\ket{0,1}_\sB = \frac{\ket{0,1}_\sxB + \ket{1,0}_\sxB}{\sqrt{2}}$
and
$\ket{1,0}_\sB = \frac{\ket{0,1}_\sxB - \ket{1,0}_\sxB}{\sqrt{2}}$,
we get
\begin{eqnarray}
\ket{1,0}_\sB \ket{F_{1,0}}_\sE &+& \ket{0,1}_\sB \ket{G_{0,1}}_\sE
\nonumber \\
&=& \frac{\ket{0,1}_\sxB}{\sqrt{2}}
\left[ \ket{G_{0,1}}_\sE + \ket{F_{1,0}}_\sE \right] \nonumber \\
&+& \frac{\ket{1,0}_\sxB}{\sqrt{2}}
\left[ \ket{G_{0,1}}_\sE - \ket{F_{1,0}}_\sE \right].
\end{eqnarray}
Since the probability of getting a photon in the
$\ket{-}$ mode must be $0$, it is necessary that
$\ket{F_{1,0}}_\sE = \ket{G_{0,1}}_\sE$.

For $m > 1$, using the ladder operators $a_\bitz$, $a_\bito$,
$a_+$, and $a_-$, since
$a_\bitz = \frac{a_+ + a_-}{\sqrt{2}}$ and
$a_\bito = \frac{a_+ - a_-}{\sqrt{2}}$,
we get
\begin{eqnarray}
\ket{0,m}_\sB &=& \frac{{a_\bitz^\dagger}^m \ket{0,0}_\sB}{\sqrt{m!}}
\nonumber \\
&=& \frac{1}{\sqrt{2^m \cdot m!}} \sum_{k = 0}^m \binom{m}{k}
{a_-^\dagger}^k {a_+^\dagger}^{m-k} \ket{0,0}_\sB \label{0_binom} \\
\ket{m,0}_\sB &=& \frac{{a_\bito^\dagger}^m \ket{0,0}_\sB}{\sqrt{m!}}
\nonumber \\
&=& \frac{1}{\sqrt{2^m \cdot m!}} \sum_{k = 0}^m \binom{m}{k}
(-1)^k {a_-^\dagger}^k {a_+^\dagger}^{m-k} \ket{0,0}_\sB.
\label{1_binom}
\end{eqnarray}
From \eqref{0_binom} and \eqref{1_binom} it follows that
\begin{eqnarray}
\ket{m,0}_\sB \ket{F_{m,0}}_\sE &+& \ket{0,m}_\sB \ket{G_{0,m}}_\sE
\nonumber \\
&=& \ket{e^{(m)}}_\sB \left[ \ket{G_{0,m}}_\sE +
\ket{F_{m,0}}_\sE\right] \nonumber \\
&+& \ket{o^{(m)}}_\sB \left[\ket{G_{0,m}}_\sE -
\ket{F_{m,0}}_\sE\right],
\label{casem}
\end{eqnarray}
with
\begin{eqnarray}
\ket{e^{(m)}}_\sB
&=& \frac{1}{\sqrt{2^m \cdot m!}} \sum_{k\text{\ even}}
\binom{m}{k} {a_-^\dagger}^k {a_+^\dagger}^{m-k} \ket{0,0}_\sB \\
\ket{o^{(m)}}_\sB &=& \frac{1}{\sqrt{2^m \cdot m!}} \sum_{k\text{\ odd}}
\binom{m}{k} {a_-^\dagger}^k {a_+^\dagger}^{m-k} \ket{0,0}_\sB,
\end{eqnarray}
where ${a_-^\dagger}^k{a_+^\dagger}^{m-k}\ket{0,0}_\sB$ is,
up to a constant factor, the Fock state $\ket{k,m-k}_\sxB$.
The probability of finding a photon in the $\ket{-}$
mode must be zero; thus, the coefficient of
${a_-^\dagger}^k {a_+^\dagger}^{m-k} \ket{0,0}_\sB$
for $k > 0$ must be zero.
Substituting $\ket{e^{(m)}}_\sB$ and $\ket{o^{(m)}}_\sB$
by their values in \eqref{casem}, the coefficient
of ${a_-^\dagger}^k{a_+^\dagger}^{m-k}\ket{0,0}_\sB$ is
(up to a non-zero constant factor)
$\ket{G_{0,m}}_\sE + \ket{F_{m,0}}_\sE$ for even $k$
and $\ket{G_{0,m}}_\sE - \ket{F_{m,0}}_\sE$ for odd $k$.
Since $k = m > 0 $ and $k = m - 1 > 0$ have different parities,
this implies both $\ket{G_{0,m}}_\sE + \ket{F_{m,0}}_\sE = 0$ and
$\ket{G_{0,m}}_\sE - \ket{F_{m,0}}_\sE = 0$, and thus
$\ket{F_{m,0}}_\sE = \ket{G_{0,m}}_\sE = 0$.
\end{proof}

\bibliography{mirror}

\end{document}